\documentclass[11pt,twoside,twocolumn,english]{article}
\usepackage{mathptmx}
\usepackage{abstract}
\usepackage[T1]{fontenc}
\usepackage[utf8]{inputenc}
\usepackage[a4paper]{geometry}
\usepackage{color}
\usepackage{babel}
\usepackage{mathrsfs}
\usepackage{amsmath}
\usepackage{amssymb}
\usepackage{esint}
\usepackage[unicode=true,pdfusetitle,
 bookmarks=true,bookmarksnumbered=false,bookmarksopen=false,
 breaklinks=false,pdfborder={0 0 1},backref=false,colorlinks=true]
 {hyperref}
\usepackage{breakurl}

\makeatletter

\@ifundefined{date}{}{\date{}}
\usepackage{babel}
\usepackage{babel}
\usepackage{babel}
\allowdisplaybreaks[4]

\makeatother

\begin{document}

\title{A scheme for determining fundamental interactions\\
 and the universality principle}
\author{Zhongmin Qian\thanks{Exeter College, Turl Street, Oxford, OX1 3DP, England. Email: zhongmin.qian@exeter.ox.ac.uk}}
\maketitle
\noindent\textbf{Summary.} The Standard Model of particle physics was established based on the
equivalence principle and gauge invariance. The Lagrangians were built
upon experimental data demonstrating the violation of discrete symmetries
together with ideas of spontaneous symmetry breaking. The features
of vector and axial-vector interactions and chirality principle are
manually added on in order to fit in the observation data, rather
than explained by the model as one may hope. Here we develop a theory
of interactions based on the Dirac algebra together with the exterior
operations the wedge product and the star operator on the space-time,
and develop a scheme for determining all possible interaction formulations.
In our scheme, the chirality transformation and the left and right
handedness come up naturally. A simple postulate about the universality
of physical interactions based on the space-time causal structure
yields the attracting features mentioned above as consequences. 

\section{Introduction}

The Standard Model (see for example \cite{Glashow1961,Salam1969,Weinberg1967,WeinbergVol2,Giunti-Kim})
of the strong, electromagnetic and weak interactions of particles
is an $SU(3)\times SU(2)\times U(1)$ gauge theory, which was built
upon the equivalence principle \cite{Einstein-general,Hawking-Ellis},
the gauge invariance principle \cite{Weyl-gauge,Yang-Mills1954,Utiyama-gauge,WeinbergVol2,Peskin-Schroeder},
and ideas of spontaneous breaking of local gauge symmetries \cite{Englert-Brout1964,Guralnik etc}
together with the Higgs mechanism \cite{Higgs1964,Guralnik etc}.
The gauge group uniquely determines the number of gauge bosons mediating
interactions with matter fields via co-variant derivatives. Lagrangians
of the Standard Model have been assembled by using experiment data,
guided mainly by the universality of weak interactions, the \emph{V-A}
current-current theory of weak interactions (the chirality principle)
\cite{Feynman-Gell-mann,Sudarshan-Marshak,Marshak etc} based on the
discoveries of the parity violation \cite{Lee-Yang1957,Wu etc} and
the \emph{CP} violation \cite{Cronin-CP,Bigi-Sanda,Branco-etc}. The
equivalence principle demands the invariance of laws of Nature under
transformations which preserve the gravitational field and the \emph{causal
structure} \cite{Hawking-Ellis}. In the context of special relativity,
the equivalence principle can be made more precise: mathematical theories
describing laws of Nature should be invariant under any change of
coordinate systems, and invariant under \emph{proper and orthochonous}
Lorentz transformations and translations of the space-time. In particular,
the parity or the time inversion conservation, which was once taken
as granted due to the correspondence principle, has been abandoned.
It is not widely appreciated that the gravitational field $(g_{\mu\nu})$
defines the \emph{duality} of various fundamental fields and its significance
in the understanding of parity and time inversion symmetries. It is
the causal structure of the space-time which is the origin of violations
of discrete symmetries.

Fermi made use of the Dirac matrices to propose a model for weak interactions,
since then the Dirac algebra has become the main ingredient in the
construction of Lagrangians within the framework of gauge theories.
The chirality matrix $\gamma^{5}$ and the helicity defined in terms
of the left and right hand projections $\frac{1}{2}(1\pm\gamma^{5})$
have been appeared manually in the Standard Model. It is a mystery
why $\gamma^{5}$ and $\frac{1}{2}(1\pm\gamma^{5})$ play so fundamental
roles in the high energy physics, and it remains to reveal their meanings.
In this article we consider the Dirac algebra combining with the exterior
calculus \cite{de Rham,Chern} on the space-time, and demonstrate
that the chirality principle and the concept of helicity are better
understood in terms of the Hodge star operator $\star$ \cite{de Rham,Chern}.
We develop a general theory of interactions which allows us to determine
Lagrangians which obey the fundamental principles of equivalence and
the gauge invariance. The violation of discrete symmetries, and the
universal theory of interactions and the chirality follow as consequences
of a simple postulate based on the space-time structure rather than
added into theoretical models manually.

\section{Universality of interactions}

Let $(g_{\mu\nu})$ denote the Minkowski metric and $\varOmega$ be
the volume form $-dx^{0}\wedge dx^{1}\wedge dx^{2}\wedge dx^{3}$.
In quantum field theory, only two kinds of fields (and their corresponding
quantized fields), tensor fields and Dirac's (fermion) fields, are
needed. The tensor fields transfer under any diffemorphism $F$ of
the space-time by the differentiation $F_{\ast}$ (see \cite{Chern}
for a definition). The exterior algebra over the space-time is equipped
with the wedge product $\wedge$ and exterior differentiation $d$.
The operations $\wedge$ and $d$ are defined on the existence of
coordinate system, and no physical processes are involved. The star
operator $\star$ is however determined by the gravitational field
$(g_{\mu\nu})$, which takes a $p$-form $\alpha=\left(\alpha_{i_{1}\cdots i_{p}}\right)$
to a $4-p$ form $\star\alpha=\left((\star\alpha)_{j_{1}<\cdots<j_{4-p}}\right)$.
$\star$ is linear operating on the exterior algebra and $\star\star=-1$,
which is introduced to determine the adjoint of $d$, $\delta=-\star d\star$.
The $\star$ operator is determined by the following rules: $\star1\rightarrow\varOmega$,
$dx^{0}\rightarrow dx^{1}\wedge dx^{2}\wedge dx^{3}$, $dx^{1}\rightarrow dx^{0}\wedge dx^{2}\wedge dx^{3}$,
$dx^{2}\rightarrow-dx^{0}\wedge dx^{1}\wedge dx^{3}$, and $dx^{3}\rightarrow dx^{0}\wedge dx^{1}\wedge dx^{2}$,
that is, a vector $A=(A_{\mu})$ is transferred to $\star A=(A_{0},A_{1},-A_{2},A_{3})$.
For two forms, the rules are given by $\star\left(dx^{0}\wedge dx^{1}\right)=-dx^{2}\wedge dx^{3}$,
$\star\left(dx^{0}\wedge dx^{2}\right)=dx^{1}\wedge dx^{3}$, $\star\left(dx^{0}\wedge dx^{3}\right)=-dx^{1}\wedge dx^{2}$,
$\star\left(dx^{1}\wedge dx^{2}\right)=dx^{0}\wedge dx^{3}$, $\star\left(dx^{1}\wedge dx^{3}\right)=-dx^{0}\wedge dx^{2}$
and $\star\left(dx^{2}\wedge dx^{3}\right)\rightarrow dx^{0}\wedge dx^{1}$.
The exterior differentiation $d$ doesn't obey the gauge invariance
principle, so that it must be replaced by gauge theories. In a gauge
theory, interactions are mediated by bosons (which are differential
forms of degree one on the space-time), which appear as connection
forms of co-variant derivatives. Suppose the gauge group is a simple
Lie group $G$ with its Lie algebra $\mathfrak{g}$. Choose a normal
basis $\left\{ T_{\alpha}\right\} $ of $\mathfrak{g}$ (in physics
literature, $t_{\alpha}=iT_{\alpha}$ are used), so that every element
in the connected component of $G$ at the identity can be expressed
as $\exp\left(\varepsilon^{\alpha}T_{\alpha}\right)$ for some reals
$\varepsilon^{\alpha}$. Consider a $\mathfrak{g}$-valued connection
$D=d+\omega$ on the space-time, where $\omega=A^{\alpha}T_{\alpha}$
and $A^{\alpha}=A_{\mu}^{\alpha}dx^{\mu}$ are bosons, real valued
differential forms on space-time. If $\varPhi=\left(\varPhi_{a}\right)$
is a family of fermion fields, then $D\varPhi=\left[D_{\mu}\varPhi\right]dx^{\mu}$,
where $D_{\mu}\varPhi=\partial_{\mu}\varPhi+A_{\mu}^{\alpha}T_{\alpha}\varPhi$.
Recall that we can only integrate differential forms of degree 4,
i.e. top forms, on the space-time, hence Lagrangians which have to
obey both the equivalence principle and the gauge invariance, must
be constructed from Lorentz invariant differential forms of degree
4 on the space-time. On the other hand, Lorentz invariant top forms
which also verify the gauge invariance must be constructed via the
bosons $A^{\alpha}T_{\alpha}$, $\mathcal{V}=\gamma_{\mu}dx^{\mu}$
together with spinors $\varPhi$ and $\varPsi$. The bosons $A^{\alpha}$
define a $\mathfrak{g}$-valued differential form of degree two $\sum_{\mu<\nu}F_{\mu\nu}^{\alpha}T_{\alpha}dx^{\mu}\wedge dx^{\nu}$,
the curvature form. There are only two possible forms of degree 4
one can build from $F$ alone, that is, $F\wedge\star F$ (which is
the gauge Lagrangian) and $F\wedge F$. The top form $F\wedge F$
is however a Chern form and therefore it has definte integration,
from which it might be possible to construct useful Lorentz and gauge
invariant quantities.

The gamma matrices were originally introduced by Dirac \cite{Dirac-gamma,Dirac3}
to derive the relativistic wave equation that $\gamma^{\mu}(i\partial_{\mu}-eA_{\mu})\psi=m\psi$,
where $\gamma=\left(\gamma^{\mu}\right)$ is a unitary representation
of Dirac matrices \cite{Ohlsson,Messiah2} so that $\gamma^{\mu}\gamma^{\nu}+\gamma^{\nu}\gamma^{\mu}=2g^{\mu\nu}$
and $\gamma^{\mu\dagger}=\gamma^{0}\gamma^{\mu}\gamma^{0}$. $\left\{ \gamma^{\mu}\right\} $
generates the Dirac algebra of 16 elements, which give rise to transformation
rules for spinors determined by von Neumann \cite{von Neumann1928}
who classified elements of the Dirac algebra into 5 classes (see Messiah
\cite{Messiah2}, page 896, Table XX.1.). The Dirac algebra was used
by Lee and Yang \cite{Lee-Yang-twocomponents}, Schwinger \cite{Schwinger1957},
Salam \cite{Salam1957}, Salam and Ward \cite{Salam-Ward}, Sudarshan
and Marshak \cite{Sudarshan-Marshak}, Feynman and Gell-mann \cite{Feynman-Gell-mann}
to generalize Fermi's theory of weak interactions into the \emph{V-A}
current-current theory \cite{kabir-weak}. The Dirac algebra has thus
become the standard tool (see \cite{Peskin-Schroeder,WeinbergVol2,Giunti-Kim}
for example) and the main ingredient in the construction of effective
Lagrangians.

Our new idea is to make use of the Dirac algebra through the exterior
calculus \cite{de Rham}.  Let $\gamma_{\mu}=g_{\mu\nu}\gamma^{\nu}$
and define a matrix-valued differential form (of degree one) $\mathcal{V}=\gamma_{\mu}dx^{\mu}$
which assigns the transformation rule as a tensor field for $\left\{ \gamma^{\mu}\right\} $
under any coordinate change of the space-time (and therefore under
every Lorentz transformation in particular). By using $\wedge$, one
can generate further three differential forms of various degrees,
namely the ``tensor'' $\mathcal{T}=\frac{1}{2}\mathcal{V}\wedge\mathcal{V}$,
the ``axial vector'' $\mathcal{A}=\frac{1}{3!}\mathcal{V}\wedge\mathcal{V}\wedge\mathcal{V}$
and the ``pseudo scalar'' $\mathcal{P}=\frac{1}{4!}\mathcal{\mathcal{V}\wedge\mathcal{V}}\wedge\mathcal{V}\wedge\mathcal{V}$,
which are the fundamental forms to construct Lorentz invariant Lagrangians.
Applying the star operator $\star$ to these fundamental forms to
generate differential forms $\star\boldsymbol{1}=\varOmega=i\gamma^{5}\mathcal{P}$,
$\star\mathcal{V}=i\gamma^{5}\mathcal{A}$, $\star\mathcal{A}=i\gamma^{5}\mathcal{V}$,
$\star\mathcal{T}=i\gamma^{5}\mathcal{T}$ and $\star\mathcal{P}=i\gamma^{5}$,
where $\gamma^{5}=i\gamma^{0}\gamma^{1}\gamma^{2}\gamma^{3}=\gamma_{5}$.
Therefore $i\gamma^{5}$ plays a role of the Hodge duality (which
is determined by the gravitational field only): 
\begin{equation}
\star\left\{ \boldsymbol{1},\mathcal{V},\mathcal{T},\mathcal{A},\mathcal{P}\right\} =i\gamma^{5}\left\{ \mathcal{P},\mathcal{A},\mathcal{T},\mathcal{V},\boldsymbol{1}\right\} .\label{eq:star1}
\end{equation}
The duality (\ref{eq:star1}) provides a clear geometric meaning of
the chirality transformation $\gamma^{5}$. Let us emphasize that
the single differential form of degree one, $\mathcal{V}$, with the
help of wedge product $\wedge$ and the star operator $\star$, generates
the Dirac algebra and determines transformation rules for all elements
of the Dirac algebra.

It is not widely recognized that Lorentz invariants can be constructed
solely through the 10 fundamental differential forms in (\ref{eq:star1})
together with operations $\wedge$ and $\star$. Let us make this
claim more explicit since we will use this feature to formulate a
fundamental postulate later on. Let us consider the current-current
theory for $\frac{1}{2}$-spin particles. Recall that spinors are
sections of the complex bundle $\mathbb{R}^{4}\times\mathbb{C}^{4}$,
and what makes them to be spinors is the action via the spin representation
of the \emph{proper orthochonous} Lorentz sub-group $\mathscr{L}_{0}$.
If $\varLambda$ is a diffeomorphism of the space-time and $\varPsi=\left(\varPsi^{\mu}\right)$
then $\varLambda_{\star}\varPsi=\varPsi\circ\varLambda^{-1}$ and
$\varLambda$ operates only on the orbital variables. If $\varLambda\in\mathscr{L}_{0}\bigcup\mathtt{P}\mathscr{L}_{0}$
(where $\mathtt{P}=\textrm{diag}(1,-1,-1,-1)$ is the space inversion),
then there are exactly two invertible matrices $S\left(\varLambda\right)$
(which differ by a sign, see \cite{W.Pauli1936,Messiah2}), such that
$\varLambda_{\;\nu}^{\mu}\gamma^{\nu}=S\left(\varLambda\right)^{-1}\gamma^{\mu}S\left(\varLambda\right)$,
$S\left(\varLambda\right)^{-1}=\gamma^{0}S\left(\varLambda\right)^{\dagger}\gamma^{0}$
and $S\left(\varLambda\right)^{\star}=B^{\dagger}S\left(\varLambda\right)B$,
where $B$ is a unitary matrix such that $\gamma^{\mu}=B\gamma^{\mu\star}B^{\dagger}$
and $BB^{\star}=B^{\star}B=-I$ (see page 900, \cite{Messiah2}).
It is direct to verify that $S(\varLambda)^{-1}\gamma^{5}S(\varLambda)$$=\det\left(\varLambda\right)\gamma^{5}$.
It is impossible to extend the spin representation to other two components
of the Lorentz group, but there is an invertible matrix $\mathtt{U}$
corresponding to the time inverse $\mathtt{T}=\textrm{diag}\left(-1,1,1,1\right)$,
such that $\mathtt{T}_{\;\nu}^{\mu}\gamma^{\nu}=U^{-1}\gamma^{\mu}U$,
$U^{\star}=B^{\dagger}UB$, however $U^{-1}=\gamma^{0}U^{\dagger}\gamma^{0}$
can not be satisfied. For Dirac or chiral representations, $S\left(\mathtt{P}\right)=\pm\gamma^{0}$
called the \emph{parity operator}, and $U=ci\gamma^{5}\gamma^{0}$
where $c$ is real. If $\varLambda\in\mathscr{L}_{0}\cup\mathtt{P}\mathscr{L}_{0}$,
$S\left(\varLambda\right)$ acts on $\varPsi$ naturally which leads
to the transformation rule $\varPsi\rightarrow S_{\varLambda}\varPsi$
given by $\varPsi\rightarrow S\left(\varLambda\right)\varPsi\circ\varLambda^{-1}$.

We may apply a differential form $\Upsilon$ in (\ref{eq:star1})
to a spinor $\varPsi$ from right and left to construct differential
forms $\Upsilon\varPsi$ and $\overline{\varPsi}\Upsilon$ respectively.
For example $\mathcal{V}\varPsi=\gamma_{\mu}\varPsi dx^{\mu}$ and
$\overline{\varPsi}\mathcal{V}=\overline{\varPsi}\gamma_{\mu}dx^{\mu}=\overline{\mathcal{V}\varPsi}$,
where $\overline{\varPsi}=\varPsi^{\dagger}\gamma^{0}$. In general
$\overline{\Upsilon\varPsi}=-\overline{\varPsi}\Upsilon$ for $\Upsilon\neq\mathcal{V}$.
These differential forms follow transformation rules under every diffeomorphism
$\varLambda$ of the space-time as tensor fields. If however $\varLambda\in\mathscr{L}_{0}\cup\mathtt{P}\mathscr{L}_{0}$,
then $\varLambda$ acts on $\mathcal{A}\varPsi$ and $\overline{\varPsi}\mathcal{A}$
by its spin representation $S\left(\varLambda\right)$ as well, and
two transformation rules may not always coincide. In fact $S_{\varLambda}\left[\varUpsilon\varPsi\right]=S(\varLambda)\varLambda_{\star}\left[\varUpsilon\varPsi\right]$
and $S_{\varLambda}\left[\overline{\varPsi}\varUpsilon\right]=\varLambda_{\star}\left[\overline{\varPsi}\varUpsilon\right]S(\varLambda)^{-1}$,
while $S_{\varLambda}\left[\star\varUpsilon\varPsi\right]=\det\left(\varLambda\right)S(\varLambda)\varLambda_{\star}\left[\star\varUpsilon\varPsi\right]$
for $\varUpsilon=I,\mathcal{V},\mathcal{T},\mathcal{A}$ or $\mathcal{P}$.
If $\det\left(\varLambda\right)=-1$, then the spin action $S_{\varLambda}$
on the field $\star\Upsilon\varPsi$ differs from the natural operation
$\varLambda_{*}$, by this we mean the parity violation (and always
to the maximum). It follows that the possible Lorentz (quadratic)
invariants are $\overline{\varPhi}\varPsi$, $\overline{\varPhi}\gamma^{5}\varPsi$.
For two pairs of spinors, in addition to quadratic products, the only
Lorentz invariants are $\left[\overline{\varPhi}\gamma_{\mu}\varPsi\right]\left[\overline{\phi}\gamma^{\mu}\psi\right]$,
$\left[\overline{\varPhi}\gamma_{5}\gamma_{\mu}\varPsi\right]\left[\overline{\phi}\gamma^{5}\gamma^{\mu}\psi\right]$,
$\left[\overline{\varPhi}\mathcal{\gamma_{\mu}}\varPsi\right]\left[\overline{\phi}\gamma^{5}\gamma^{\mu}\psi\right]$,
also the following couplings which will not appear in any pair production:
$\left[\overline{\varPhi}\gamma_{\mu}\gamma_{\nu}\varPsi\right]\left[\overline{\phi}\gamma^{\mu}\gamma^{\nu}\psi\right]$,
$\left[\overline{\varPhi}\gamma_{\mu}\gamma_{\nu}\varPsi\right]\left[\overline{\phi}\gamma^{5}\gamma^{\mu}\gamma^{\nu}\psi\right]$
and $\left[\overline{\varPhi}\gamma_{5}\gamma_{\mu}\gamma_{\nu}\varPsi\right]\left[\overline{\phi}\gamma^{5}\gamma^{\mu}\gamma^{\nu}\psi\right]$
(where summations run over $\mu<\nu$).

It remains to explain the mystery why the operators $1\pm\gamma^{5}$
enter into the construction of the standard model? To answer this
question, we observe that there are exactly two differential forms
of degree one, namely $\mathcal{V}=\gamma_{\mu}dx^{\mu}$ and $-i\star\mathcal{A}=\gamma_{5}\gamma_{\mu}dx^{\mu}$.
Nature may select the simplest possible mechanism to mediate interactions
among the most fundamental particles to reveal her beauty of laws.
With this faith, and the observation that all interactions appear
in gauge theories are mediated and coupled with fields and the differential
form $\mathcal{V}$ of degree one via $\wedge$ and $\star$ , one
naturally asks the question why $\mathcal{V}$? To preserve the Lorentz
invariance, we can start with a general differential form of degree
one and we would like to derive natural conditions which must be independent
of any interactions or fields, however ensure the \emph{simplicity}
of laws of Nature. The most general differential form of degree one
can be written as $\mathcal{F}=\lambda_{1}\mathcal{V}-\lambda_{2}i\star\mathcal{A}$.
We seek for an $\mathcal{F}$ which generates a simpler algebra as
possible (by using $\wedge$ and $\star$). To this end we calculate
that 
\begin{equation}
\mathcal{F}\wedge\mathcal{F}=\left(\lambda_{1}^{2}-\lambda_{2}^{2}\right)\sum_{\mu<\nu}\gamma_{\mu}\gamma_{\nu}dx^{\mu}\wedge dx^{\nu},\label{eq:ff}
\end{equation}
\begin{align}
\mathcal{F}\wedge\mathcal{F}\wedge\mathcal{F} & =3\left(\lambda_{1}^{2}-\lambda_{2}^{2}\right)\left(\lambda_{1}+\lambda_{2}\gamma^{5}\right)\label{eq:fff}\\
 & \sum_{\sigma<\mu<\nu}\gamma_{\sigma}\gamma_{\mu}\gamma_{\nu}dx^{\sigma}\wedge dx^{\mu}\wedge dx^{\nu}
\end{align}
and 
\begin{equation}
\mathcal{F}\wedge\mathcal{F}\wedge\mathcal{F}\wedge\mathcal{F}=-12\left(\lambda_{1}^{2}-\lambda_{2}^{2}\right)^{2}i\gamma^{5}\varOmega.\label{eq:ffff}
\end{equation}
It follows that the simplest algebra generated by $\mathcal{F}$ is
achieved when $\mathcal{F}\wedge\mathcal{F}=0$, i.e. the differential
one form $\mathcal{F}$ is self-dual. This self-duality of $\mathcal{F}$
is equivalent to that $\lambda_{1}=\pm\lambda_{2}$, which leads to
two possible differential forms (up to a multiple constant) $\mathcal{L}=\frac{1}{2}\left(1+\gamma^{5}\right)\gamma_{\mu}dx^{\mu}$
and $\mathcal{R}=\frac{1}{2}\left(1-\gamma^{5}\right)\gamma_{\mu}dx^{\mu}$,
and in this way the projections $1\pm\gamma^{5}$ appear naturally.
One may work out the algebra generated by $\mathcal{L}$ and $\mathcal{R}$
too through the exterior multiplication $\wedge$ and the star operator
$\star$. The results are $\mathcal{L}\wedge\mathcal{R}=(1+\gamma^{5})\mathcal{T}$,
$\mathcal{R}\wedge\mathcal{L}=(1-\gamma^{5})\mathcal{T}$, $\mathcal{L}\wedge\mathcal{R}\wedge\mathcal{L}=3(1+\gamma^{5})\mathcal{A}$
and $\mathcal{R}\wedge\mathcal{L}\wedge\mathcal{R}=3(1-\gamma^{5})\mathcal{A}$.
It follows that there are only 6 (instead of 10 for Dirac algebra)
non-trivial differential forms one may constructed from $\mathcal{L}$
and $\mathcal{R}$, two for each degree. The six elements are self-dual
in the sense that $\star\left(\mathcal{L}\wedge\mathcal{R}\right)=i\mathcal{L}\wedge\mathcal{R}$,
$\star\left(\mathcal{R}\wedge\mathcal{L}\right)=-i\mathcal{R}\wedge\mathcal{L}$,
$\star\mathcal{L}=\frac{i}{3!}\mathcal{L}\wedge\mathcal{R}\wedge\mathcal{L}$
and $\star\mathcal{R}=-\frac{i}{3!}\mathcal{R}\wedge\mathcal{L}\wedge\mathcal{R}$.
In particular, the differential form $\mathcal{L}$, in contrast with
the case of $\mathcal{V}$, only generates one more differential form
(of degree three) $\star\mathcal{L}$. Observing the amazing
duality, one can not stop to formulate the following proposition.

\emph{Postulate of Universality}. The weak interactions are mediated
through $\mathcal{L}$ (or equivalently $\mathcal{R}$) only.

Under this version of the universality principle, the only coupling
allowed under this postulate is the top form $\left[\overline{\varPhi}\mathcal{L}\varPsi\right]\wedge\left[\overline{\phi}\star\mathcal{L}\psi\right]$,
which has the density (up to a constant) $\left[\overline{\varPhi}(1+\gamma^{5})\gamma_{\mu}\varPsi\right]\left[\overline{\phi}(1+\gamma^{5})\gamma^{\mu}\psi\right]$.
Therefore the universal \emph{V-A} theory for weak interactions follows.
There is only one possible coupling with bosons and spinors one can
build, namely $\overline{\varPhi}\star\mathcal{L}\wedge D\varPsi$
which gives the interaction $\overline{\varPhi}(1+\gamma^{5})\left[\gamma^{\mu}D_{\mu}\varPsi\right]$
up to a constant. This is exactly the chirality principle. While,
there is no non-trivial two form generated by $\mathcal{L}$ alone,
therefore, a kinetic term involving a quadratic of co-variant derivatives
of matter fields such as $\sum_{\mu\neq\nu}\overline{\left[\gamma^{\mu}D_{\mu}\varPhi\right]}\left[\gamma^{\nu}D_{\nu}\varPsi\right]$,
which is Lorentz and gauge invariant, is forbidden.

In order to cover other interactions, we may extend the universality
principle.

\emph{Postulate of General Universality}. The physical interactions are mediated by $\mathcal{L}$, $\mathcal{R}$, $\mathcal{L}\wedge\mathcal{R}$ and $\mathcal{R}\wedge\mathcal{L}$, which can not be mixed. Therefore 4 internal numbers can be assigned for interactions.

The postulate needs to be verified in terms of experimental data
which we will not do here, and leave to the more competent experts
in particle physics. We would like to give some consequences derived
from this postulate. There are only two possible interactions describing
the interactions with matter fields mediated by bosons, namely the
top forms $\overline{\varPhi}\star\mathcal{R}\wedge D\varPsi$ and
$\overline{\varPhi}\star\mathcal{L}\wedge D\varPsi$ which give rise
to the densities $\overline{\varPhi}\left(1\pm\gamma^{5}\right)\gamma^{\mu}D_{\mu}\varPsi$.
Moreover kinetic terms are not forbidden, and in fact there are exactly
two top forms $\overline{D\varPhi}\wedge\mathcal{L}\wedge\mathcal{R}\wedge D\varPsi$
and $\overline{D\varPhi}\wedge\mathcal{R}\wedge\mathcal{L}\wedge D\varPsi$,
both are Lorentz and gauge invariant, and therefore there are two
possible kinetic terms $\sum_{\mu<\nu}\overline{\gamma^{\mu}D_{\mu}\varPhi}\left(1\pm\gamma^{5}\right)\gamma^{\nu}D_{\nu}\varPsi$
which can be embedded into Lagrangians for interactions.

{\small{}{ }{\small \par}

{\small{}}}{\small \par}
\end{document}